\documentclass[aps,prl,twocolumn,showpacs,amsmath,epsfig,amssym,letterpaper]{revtex4}
\usepackage[dvips]{graphicx}
\begin{document}
\date{January 2002}
\title{Qubitless Quantum Logic}
\author{Richard B. Kassman$^{1,2}$, Gennady P. Berman$^1$, Vladimir I.
  Tsifrinovich$^3$, and Gustavo V. L\'opez$^4$\\ \ \\
  \small{$^1$Theoretical Division and CNLS, Los Alamos National
    Laboratory, Los Alamos, New Mexico, 87544.\\ $^2$Department of
    Physics, University of Illinois at Urbana-Champaign, 1110 West
    Green Street, Urbana, Illinois, 61801. \\ $^3$IDS Department,
    Polytechnic University, Six Metrotech Center, Brooklyn, New York,
    11201. \\
    $^4$Departamento de F\'isica, Universidad de Guadalajara,
    Corregidora 500, SR 44420, Guadalajara, Jalisco, Mexico. \\ \ \\}}
\begin{abstract}
  We discuss the implementation of quantum logic in a system of
  strongly interacting particles.  The implementation is qubitless since
  ``logical qubits'' don't correspond to any physical two-state
  subsystems.  As an illustration, we present the results of
  simulations of the quantum controlled-NOT gate and Shor's algorithm
  for a chain of spin-1/2 particles with Heisenberg coupling.  Our
  proposal extends the current theory of quantum information processing
  to include systems with permanent strong coupling between the two-state
  subsystems.
\end{abstract}
 \maketitle
\section{Introduction}
Current quantum computation and quantum information processing
theory is based on the manipulation of two-state
subsystems known as physical qubits \cite{1}-\cite{3}. (The
physical qubits need not be identical to the binary digits used
for encoding information, which are called logical qubits). One
begins with a system of weakly interacting or noninteracting
physical qubits and, in order to implement quantum logic gates,
one either applies short pulses of an external field or ``turns
on'' strong coupling between the qubits for short periods of time.
For example, in a recent proposal \cite{a}, the logical qubit
states $|0\rangle$ and $|1\rangle$ are represented by selected
states of three spins in a chain. The states used have $I=1/2$ and
$m_I=1/2$, thus,
\begin{eqnarray}
  &|0\rangle={1\over\sqrt2}(|\uparrow\downarrow\rangle-
  |\downarrow\uparrow\rangle)|\uparrow\rangle,&
  \nonumber \\
  &|1\rangle=\sqrt{2\over3}|\uparrow\uparrow\downarrow
  \rangle-{1\over\sqrt6}(|\uparrow\downarrow\rangle+|\downarrow\uparrow\rangle)
  |\uparrow\rangle.&
\end{eqnarray}
Quantum logic gates are implemented in this proposal by turning on
the Heisenberg interaction between neighboring spins for short
time intervals.

Physical systems with permanent strong interactions between the
subsystems are not included in the current theory of quantum
computation.  We believe that there is a need to extend the theory
to include these systems because such interactions can quite
realistically be expected in many of the types of physical systems
that have been proposed as quantum computers.  In this paper, we
discuss the implementation of quantum information processing in
these systems.  As an example, we consider a spin-1/2 Heisenberg
chain in a non-uniform magnetic field.  We assume that the
Heisenberg coupling cannot be turned off, and thus it is
impossible to separate the system into weakly interacting
subsystems. Nevertheless we show that qubitless quantum logic may
be implemented using external electromagnetic pulses.  Although
this procedure is not scalable to quantum computers with large
numbers of qubits, these systems nevertheless can be of use in
small quantum devices for opto-electronic interfaces, quantum
cryptography and communication, etc.  These small devices would
store quantum information and perform some elementary quantum
manipulation.

\section{Logical Qubits without Physical Qubits}

We now discuss general requirements for the implementation of
quantum logic in a many-level system.  We use each energy level to
represent a number.  For example, the numbers 0 to 3 can be
represented within a four-level system.  Each stationary state of
the system is identified with a binary number,
\begin{equation} \label{eq:2qstates}
  |00\rangle_l,\qquad |01\rangle_l,\qquad |10\rangle_l,\qquad
 \textrm{or}\qquad |11\rangle_l.
\end{equation}
The subscript $l$ denotes that this is a logical notation and the
qubits here should not be identified with any physical entities.
Since any quantum logic gate can be decomposed into a sequence of
one-qubit rotations and two-qubit controlled-NOT
\newcommand{\CN}{\mathsf{CN}}($\CN$) gates \cite{b}, we would like to
be able to implement these gates on the logical qubits.  For
instance, a ``$\pi$/2''-rotation \cite{2} on logical qubit ``1'',
\newcommand{\U}{\mathsf{U}}$\U_1\equiv\U_1(\pi/2)$, corresponds to
the following set of transformations, {\setlength\arraycolsep{2pt}
  \begin{eqnarray}\label{eq:u1}
  \U_1|00\rangle_l&=&{1\over\sqrt2}(|00\rangle_l+i|10\rangle_l),
  \nonumber \\
  \U_1|01\rangle_l&=&{1\over\sqrt2}(|01\rangle_l+i|11\rangle_l),
  \nonumber \\
  \U_1|10\rangle_l&=&{1\over\sqrt2}(|10\rangle_l+i|00\rangle_l),
  \nonumber \\
  \U_1|11\rangle_l&=&{1\over\sqrt2}(|11\rangle_l+i|01\rangle_l),
\end{eqnarray}}
and the modified controlled-NOT gate, $\CN_{10}$, would operate in
the following manner, {\setlength\arraycolsep{2pt}
\begin{eqnarray}\label{eq:cn}
  \CN_{10}|00\rangle_l&=&|00\rangle_l,
  \nonumber \\
  \CN_{10}|01\rangle_l&=&|01\rangle_l,
  \nonumber \\
  \CN_{10}|10\rangle_l&=&i|11\rangle_l,
  \nonumber \\
  \CN_{10}|11\rangle_l&=&i|10\rangle_l.
\end{eqnarray}}

Now, we assume that we can drive transitions between the energy
levels by applying a cyclic potential, $V$, whose frequency is
resonant to the energy difference between the levels.  A
transition between two states is possible only if the matrix
element of $V$ between those states is nonzero.  We assume that
any transition which ``flips'' one of the logical qubits is
allowed.  This assumption is valid for the model discussed below.
If a complete transition between two states takes a time $T$ to
complete, then application of $V$ for this period of time is known
as a $\pi$-pulse.  If a pulse is applied to one of the eigenstates
for a time interval $T$/2 ($\pi$/2-pulse), then a uniform
superposition of two states will be created (assuming only
resonant transitions occur).  For example, upon application of a
$\pi$/2-pulse to the state $|00\rangle_l$, resonant to the energy
difference between $|00\rangle_l$ and $|10\rangle_l$, the
transformation which results is,
\begin{equation}
  |00\rangle_l\rightarrow{1\over\sqrt2}(|00\rangle_l+i|10\rangle_l),
\end{equation}
and if the same pulse is applied to $|10\rangle_l$, the
transformation is,
\begin{equation}
  |10\rangle_l\rightarrow {1\over\sqrt2}(|10\rangle_l+i|00\rangle_l).
\end{equation}

These are the first and third transformations in (\ref{eq:u1}).
The second and fourth transformations can be implemented by the
application of a similar pulse, resonant to the energy difference
between $|01\rangle_l$ and $|11\rangle_l$.  Thus, the one-qubit
rotation, $\U_1$, can be implemented by two resonant
$\pi$/2-pulses. The $\CN_{10}$ gate (\ref{eq:cn}), on the other
hand, can be realized by the application of one $\pi$-pulse,
resonant to the energy difference
$|10\rangle_l\leftrightarrow|11\rangle_l$.  These are idealized
implementations of the gates, as in reality non-resonant
transitions occur, which may give rise to deviations from the
ideal output.

\section{Heisenberg Chain Quantum Computer}
\label{sec:hchain} Consider a chain of $L$ spins with uniform
nearest-neighbor Heisenberg coupling, $J$.  The system is
subjected to a number of radio-frequency magnetic pulses. The
Hamiltonian of the system under the influence of the $j$th pulse,
(with $\hbar=1$) is,
\begin{gather}
  {\cal H}=H+V,
  \nonumber \\
  H=-\sum_{k=0}^{L-1}\omega_k I^z_k-2J\sum_{k=0}^{L-2} \mathbf{I}_k
  \cdot \mathbf{I}_{k+1},
  \nonumber \\
  V=-{\Omega\over2}\sum_{k=0}^{L-1}
  \left(e^{-i(\nu_jt+\phi_j)}I^-_k+e^{+i(\nu_j t+\phi_j)}I^+_k\right),
\end{gather}
where $\omega_k$ is the Larmor frequency of the $k$th spin,
$\mathbf{I}_k$ is the spin operator for the $k$th spin, $\Omega$
is the Rabi frequency and $\nu_j$ and $\phi_j$ are the frequency
and phase of the pulse.  Upon diagonalization of $H$, we obtain
$2^L$ eigenstates.  Except for two of these states (the ones with
spin projection quantum numbers $m_I=\pm L/2$), they are entangled
superpositions of the individual spin product states.

In the case of uniform Larmor frequency, both the total spin and
spin projection operators, $I^2_{tot}$ and $I^z_{tot}$, commute
with $H$. Thus the eigenstates have definite values of the total
spin quantum number, $I$, and the spin projection quantum number,
$m_I$. The effect of the application of a pulse, assuming only
resonant transitions occur, is to give rise to transitions which
change the spin projection by $\Delta m_I=\pm1$, but leave the
total spin unchanged.  Starting from the ground state, which has
$I=L/2$, only $2I+1=L+1$ levels can be accessed using successive
resonant transitions resulting from the application of pulses, out
of the total number of $2^L$ levels.

Thus, in order to be able to access all possible levels, a
non-uniform Larmor frequency (i.e., a non-uniform external
magnetic field) is required.  In this case, $H$ commutes with
$I^z_{tot}$ but not with $I^2_{tot}$.  Thus, the eigenstates have
definite values of $m_I$ but one is not restricted to the subspace
of states with a fixed total spin.

\section{Controlled-NOT Gate}
\label{sec:cnot}

As a simple example of the implementation of quantum logic in the
Heisenberg system, we consider the modified controlled-NOT gate,
$\CN_{10}$, in a two-spin system.  We label the eigenstates of $H$
(expressed in terms of the eigenstates of $I^z_1$ and $I^z_2$)
with binary numbers, thus,
\begin{eqnarray}
  &|00\rangle_l=|\uparrow\uparrow\rangle,&
  \nonumber \\
  &|01\rangle_l=\alpha_1|\uparrow\downarrow\rangle+\alpha_2|\downarrow
  \uparrow\rangle,&
  \nonumber \\
  &|10\rangle_l=-\alpha_2|\uparrow\downarrow\rangle+\alpha_1|\downarrow
  \uparrow\rangle,&
  \nonumber \\
  &|11\rangle_l=|\downarrow\downarrow\rangle,&
\end{eqnarray}
Where $\alpha_1$ and $\alpha_2$ are functions of the ratio
$J/\delta\omega$, where $\delta\omega=\omega_1-\omega_0$.  In the
limit $J/\delta\omega\rightarrow\infty$,
$\alpha_1=\alpha_2=1/\sqrt2$, and in the limit
$J/\delta\omega\rightarrow0$, $\alpha_1=1$ and $\alpha_2=0$.  Note
that in this scheme, not only are the logical states entangled
superpositions of the spins, but the ``0'' and ``1'' states of the
logical qubits can't even be identified with any well-defined
states of the overall system.  Thus the scheme is qubitless in the
sense that a logical qubit doesn't correspond to any sort of
physical subsystem.

The energies of the eigenstates are given by the following
expressions:
\begin{itemize}
\item []for the states $|00\rangle_l$ and $|11\rangle_l$ respectively,
  \begin{equation}
    E_{0,3}=-{J\over2}\mp{{\omega_0+\omega_1}\over2},
  \end{equation}
\item []and for the states $|01\rangle_l$ and $|10\rangle_l$
  respectively,
  \begin{equation}
    E_{1,2}=+{J\over2}\mp{\delta\omega\over2}.
  \end{equation}
\end{itemize}
The non-zero matrix elements of the pulse, $V$, in this system
are, {\setlength\arraycolsep{2pt}\begin{eqnarray}
  _l\langle01|V|00\rangle_l=&
  _l\langle11|V|01\rangle_l=&-{\Omega_0\over2}e^{-i\nu t-\phi},\nonumber\\
  _l\langle10|V|00\rangle_l=&
  _l\langle11|V|10\rangle_l=&-{\Omega_1\over2}e^{-i\nu t-\phi},
\end{eqnarray}}
where $\Omega_0$ and $\Omega_1$ are ``effective'' Rabi
frequencies,
\begin{equation}
  \Omega_{0,1}=(\alpha_1\pm\alpha_2)\Omega.
\end{equation}
Therefore, all transitions involving a flip of one of the logical
qubits are allowed.

Using the Schr\"odinger equation, we can write equations of motion
for the probability amplitudes of the logical states, $C_{00}$,
$C_{01}$, etc.  In the interaction representation, the evolution
of each of the amplitudes will be coupled to the amplitudes of two
other states.  For example, the equation of motion for $C_{11}$
is,
\begin{equation}
  i\dot C_{11}=-{\Omega_0\over2}e^{i[(E_3-E_1-\nu)t-\phi]}C_{01}- {\Omega_1 \over2}
  e^{i[(E_3-E_2-\nu)t-\phi]}C_{10}.
\end{equation}
If we apply a pulse resonant to the frequency of the transition
$|10\rangle_l\leftrightarrow|11\rangle_l$, that is $\nu=E_3-E_2$
then, ignoring the non-resonant transitions, the solution for
$C_{11}$ is,
\begin{equation}
  C_{11}(t)=C_{11}(0)\cos\left({\Omega_1t \over 2
      }\right)+ie^{-i\phi}C_{10}(0)\sin\left({\Omega_1t \over2}\right).
\end{equation}
Similarly, the solution for $C_{10}$ with this pulse is,
\begin{equation}
  C_{10}(t)=C_{10}(0)\cos\left({\Omega_1t \over
  2}\right)+ie^{+i\phi}C_{11}(0)\sin\left({\Omega_1t \over2}\right).
\end{equation}
Thus, a pulse with a duration $\pi$/$\Omega_1$ (which is known as
a $\pi$-pulse), causes a complete exchange of probabilities
between $|10\rangle_l$ and $|11\rangle_l$, while one with a
duration $\pi/2\Omega_1$ ($\pi$/2-pulse) causes $|10\rangle_l$ to
be transformed into a uniform superposition of the two states and
likewise for $|11\rangle_l$.  The $\pi$-pulse of frequency
$\nu=E_3-E_2$ and phase $\phi=0$ is clearly what's required to
implement the modified $\CN_{10}$-gate (\ref{eq:cn}).

\section{Errors in the Implementation of the Controlled-NOT Gate}
\label{sec:cnerror}

In our preceding discussion, we assumed that pulses only drive
resonant transitions.  However, non-resonant transitions do occur
with a finite probability, which can give rise to significant
errors even in a single-pulse implementation of a quantum gate.
Here we present the results of numerical simulations of the
implementation of $\CN_{10}$. As an example, we begin with the
following initial amplitudes for the logical states,
{\setlength\arraycolsep{2pt}
\begin{eqnarray}
  &C_{00}(0)={1\over\sqrt2},\quad C_{01}(0)=0,\quad
  C_{10}(0)={1\over\sqrt2},\quad \textrm{and}&\nonumber\\
  &C_{11}(0)=0,&
\end{eqnarray}}
and present the simulated final probabilities after a pulse of
duration $\tau=\pi/\Omega_1$. The ideal output amplitudes are,
{\setlength\arraycolsep{2pt}
\begin{eqnarray}
  &C_{00}(0)={1\over\sqrt2},\quad C_{01}(0)=0,\quad
  C_{10}(0)=0,\quad \textrm{and}&\nonumber\\
  &C_{11}(0)={i\over\sqrt2}.&
\end{eqnarray}}
\begin{figure*}[bt] \begin{center}
\includegraphics[scale=0.65,angle=270]{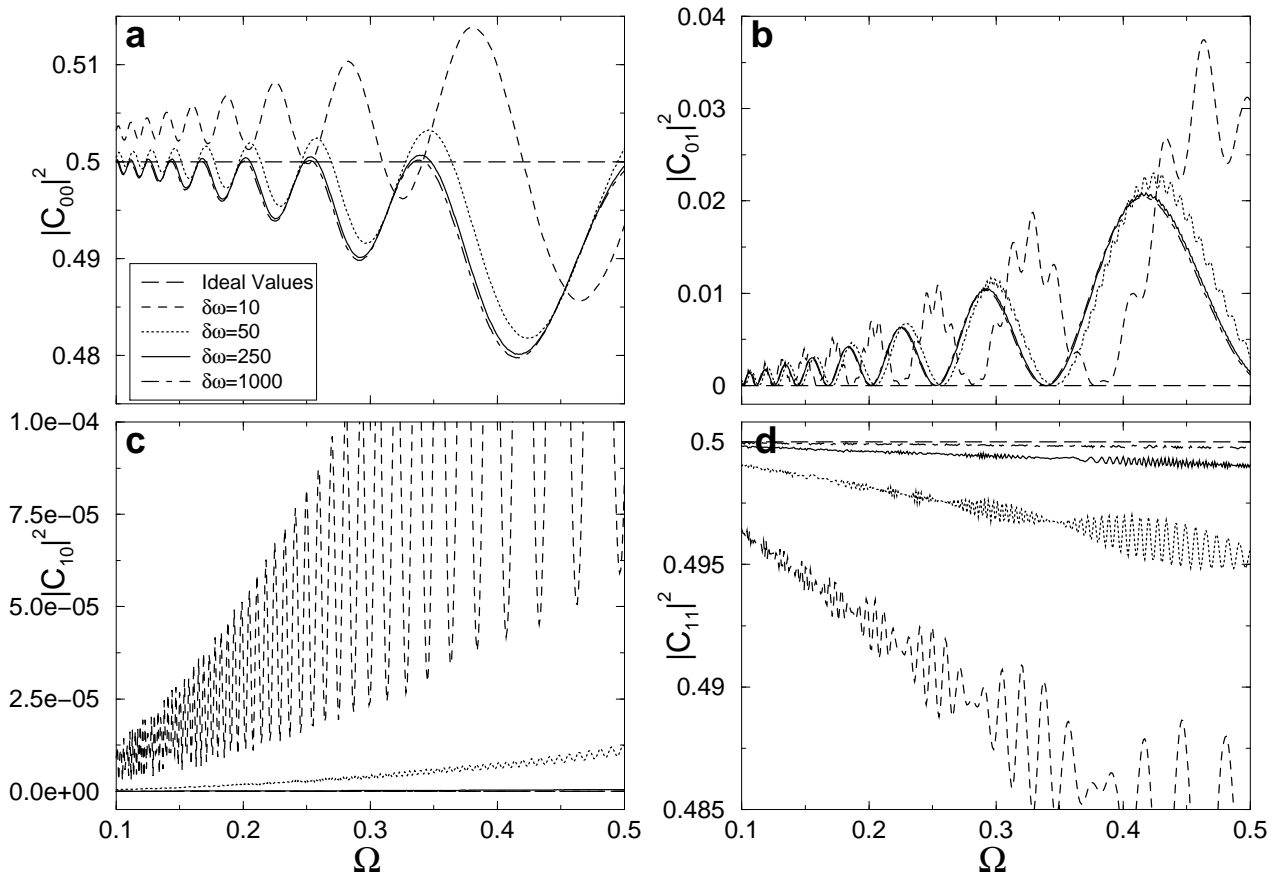}
\end{center}\caption{\label{fig:cn}Dependence of final $\CN_{10}$-gate probabilities,
  $|C_{kl}|^2$ on the Rabi frequency, $\Omega$, for different values
  of $\delta \omega$.  The values of $\Omega$ and $\delta \omega$ are
  given in units of $J$. (a) $|C_{00}|^2$. (b) $|C_{01}|^2$. (c)
  $|C_{10}|^2$. (d) $|C_{11}|^2$.}
\end{figure*}

The simulated final probabilities are plotted in Fig. \ref{fig:cn}
as functions of the Rabi frequency for four different values of
$\delta\omega$.  All parameters in the simulation are given in
units of $J$.  We chose $\omega_0=100$. One can see that the
variation of the probabilities definitely shows a dependence on
$\delta\omega$ as well as an oscillatory dependence on the Rabi
frequency.  These dependencies can be explained by dividing the
non-resonant transitions into two categories:
\begin{itemize}
\item {\it Near-Resonant Transitions}.  In this case there is only
  one such transition, namely
  $|00\rangle_l\leftrightarrow|01\rangle_l$ which, for
  $\delta\omega=50$, has a transition frequency of
  $\nu=E_1-E_0=99.98$ as compared to $\nu=E_3-E_2=98.98$ for the
  resonant transition.
\item {\it Off-Resonant Transitions}.  In this case there are two,
  $|00\rangle_l\leftrightarrow|10\rangle_l$ and
  $|01\rangle_l\leftrightarrow|11\rangle_l$ which, for
  $\delta\omega=50$, have transition frequencies of 151.02 and 149.02
  respectively.
\end{itemize}
The effects of all of the above transitions can clearly be seen in
the $\delta\omega=10$ case.  In Fig. \ref{fig:cn}(b), which shows
the final values of $|C_{01}|^2$, the large amplitude low
frequency oscillations are the result of the near-resonant
transitions and the small amplitude high frequency oscillations
are the result of the off-resonant transitions.  In Fig.
\ref{fig:cn}(c) for $|C_{10}|^2$, the primary effect is that of
the high frequency oscillations since this state is not directly
affected by the near-resonant transition, rather it is affected by
an off-resonant one. Figure \ref{fig:cn}(d) for $|C_{11}|^2$ again
shows the high frequency oscillations. However, the oscillations
are modulated and shifted by the near-resonant transition. This
may be explained by a two-step process,
$|11\rangle_l\rightarrow|01\rangle_l \rightarrow|00\rangle_l$, an
off-resonant transition followed by a near-resonant one.  In Fig.
\ref{fig:cn}(a) for $|C_{00}|^2$ the probability is primarily
influenced by the near-resonant transition, but it is shifted
upwards due to the off-resonant ones.

In order to decrease the errors on the $\CN_{10}$, one can reduce
the Rabi frequency, which reduces the effect of all non-resonant
transitions.  Another method of error reduction can be explained
by an analysis of Fig. \ref{fig:cn}.  The graphs clearly show
that increasing $\delta\omega$ decreases the influence of the
off-resonant transitions.  This is due to the fact that it
increases the difference between the frequencies of the
off-resonant transitions on the one hand, and that of the resonant
one on the other.  When $\delta\omega$ is 250, the effect of the
off-resonant transitions is almost negligible.

The remaining error is due to the near-resonant transition.  This
error can be reduced by making use of the ``$2\pi k$-method''
\cite{2,c}, which essentially involves choosing values of the Rabi
frequency where the error due to the near-resonant transition is
zero. That is, a Rabi frequency is chosen such that a $\pi$-pulse,
while giving rise to the resonant transition will also cause an
angular change of $2\pi k$ (where $k$ is an integer), due to the
near-resonant one; so that the probabilities of the states
affected by the near-resonant transition return to their initial
values at the end of the pulse.  The condition for this is,
\begin{equation}
  \sqrt{\Omega_0^2+\Delta^2} \,\tau = 2\pi k,
\end{equation}
where $\tau=\pi/\Omega_1$ as before, and $\Delta$ is the
difference between the near-resonant and resonant transition
frequencies.  Thus, for $\delta\omega\gtrsim 250$, the errors can
effectively be reduced by choosing values for the Rabi frequency
satisfying the $2\pi k$ condition.  (Here, we don't consider more
sophisticated pulse shaping methods, which can also be used to
reduce the errors in the implementation of quantum logic gates
\cite{e}).

\section{Shor's Algorithm}
\begin{figure*}[bt] \begin{center}
\includegraphics[scale=0.7,angle=270]{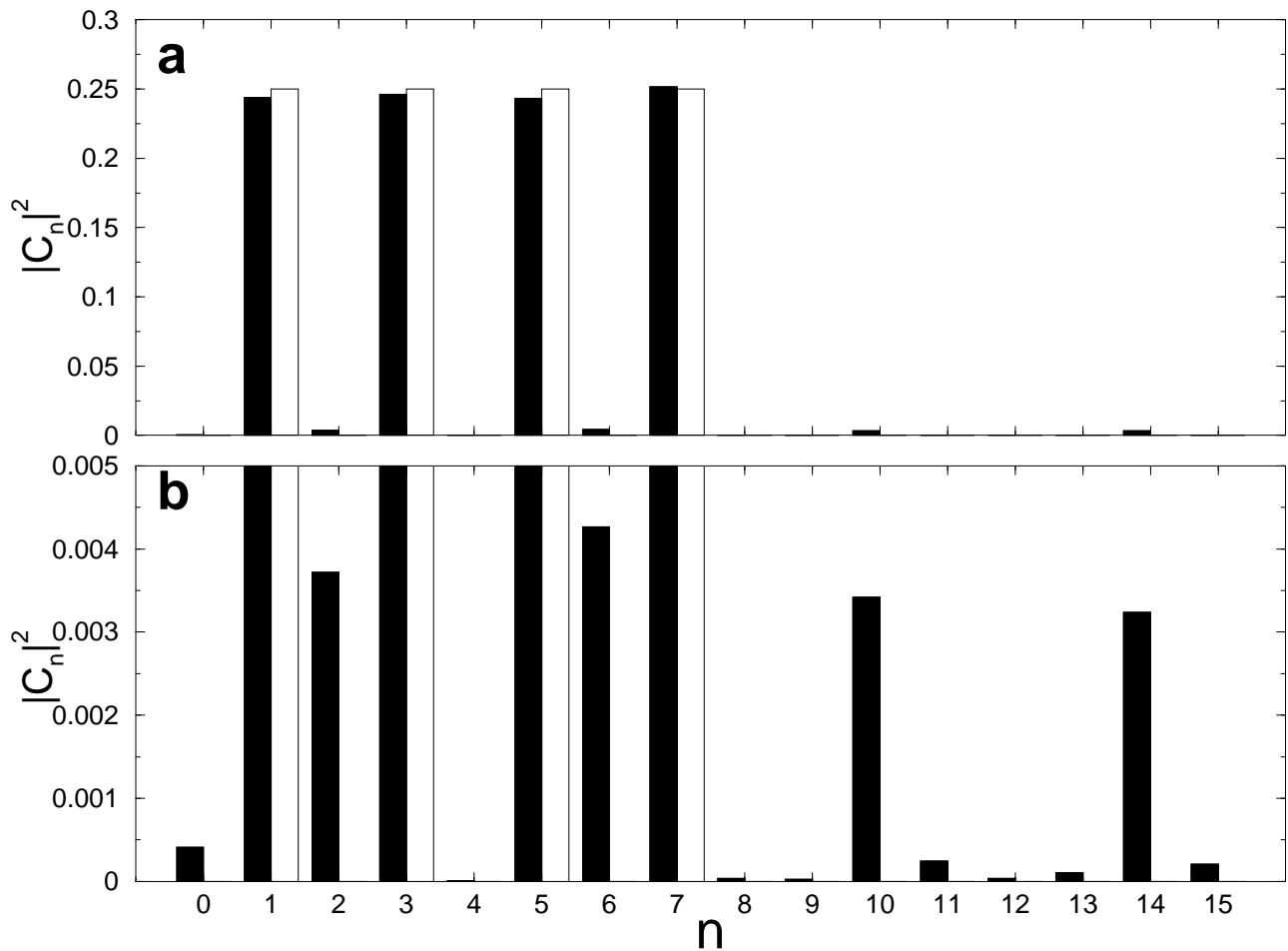}
\end{center}\caption{\label{fig:shor}(a) Probabilities of the
  logical states, $|C_n|^2$, following the
  41-pulse sequence for Shor's algorithm.  The values of the
  parameters used in our simulations were, $J=30$, $\omega_0=100$,
  $\delta \omega=30$, $\Omega=0.5$. The white bars indicate ideal
  output probabilities of Shor's algorithm, and the black bars are
  the results of our simulation. (b) The same probabilities, on an
  enlarged scale, to show the probabilities of unwanted logical
  states.}
\end{figure*}
\label{sec:shor} In order to illustrate the implementation of a
quantum algorithm in our system, we have performed a simulation of
Shor's algorithm for prime factorization\cite{d}.  We simulated a
pulse sequence for the factorization of the number four in a
four-spin Heisenberg chain quantum computer.  This is a 16
level-system and the logical states are $|0000\rangle_l,\
|0001\rangle_l,\ |0010\rangle_l,\ \dots\ |1111\rangle_l$, or in
decimal notation $|0\rangle_l,\ |1\rangle_l,\ |2\rangle_l,\ \dots\
|15\rangle_l$.  We used the same sequence of quantum gates as the
simulations on the Ising chain quantum computer \cite{c}, but the
pulses used were those appropriate to the Heisenberg case.

Shor's algorithm requires two registers of qubits.  We designate
the left two logical qubits as the x-register and the right as the
y-register.  We assume that the initial state is the ground state
$|0000\rangle_l =|0\rangle_l$.  The algorithm then proceeds in
three steps.  The first step is to create a uniform superposition
in the logical x-register.  The state which results from this is,
{\setlength\arraycolsep{2pt}
\begin{eqnarray}
|\Psi\rangle&=&{1\over2}\,(|0000\rangle_l+|0100\rangle_l+
|1000\rangle_l+|1100\rangle_l)\nonumber \\
&\equiv&{1\over2}\,(|0\rangle_l+|4\rangle_l+|8\rangle_l+|12\rangle_l),
\end{eqnarray}
The resonant pulse sequence to create this state consists of three
$\pi/2$-pulses - for the transitions $|0\rangle_l \leftrightarrow
|4\rangle_l$, then $|0\rangle_l \leftrightarrow |8\rangle_l$, then
$|4\rangle_l \leftrightarrow |12\rangle_l$.  The next step is to
transform to the state,
\begin{equation}
|\Psi\rangle= {1\over2} \sum_{x=0}^3|x,\,y(x)\rangle_l,
\end{equation}
where $y(x)=3^x$(mod 4). This is implemented using six
$\pi$-pulses. The final step is to perform a discrete Fourier
transform on the logical x-register.  This can be implemented
using a sequence of so-called $\mathsf A$- and $\mathsf B$-gates,
which in our system requires 32 ($\pi$- and $\pi/2$-) pulses. The
final state after implementation of Shor's algorithm ideally
contains the following four states with equal probability,
\begin{displaymath}
|0001\rangle_l,\qquad|0011\rangle_l,\qquad|0101\rangle_l,\qquad\textrm{and}\qquad|0111\rangle_l,
\end{displaymath}
or,
\begin{equation}
\label{eq:shorf}
|1\rangle_l,\qquad|3\rangle_l,\qquad|5\rangle_l,\qquad\textrm{and}\qquad|7\rangle_l.
\end{equation}
Figure \ref{fig:shor} shows the results of our numerical
simulations of the 41-pulse sequence.  The white bars correspond
to the ideal values of the probabilities (1/4) of the states in
(\ref{eq:shorf}).  The black bars are the probabilities resulting
from numerical simulations of the pulse sequence.  The larger
probabilities are shown in Fig. \ref{fig:shor}(a), and the smaller
ones corresponding to errors are shown on an enlarged scale in
Fig. \ref{fig:shor}(b).  It can be seen that the simulated
probabilities for the states in (\ref{eq:shorf}) are very close to
their ideal values. The maximum deviation on these states is
0.0066.  The probabilities of the unwanted states do not exceed
0.0043.  Four unwanted states have probabilities between 0.0030
and 0.0043, and the remainder have probabilities smaller than
0.0005.  The sum of the probabilities of the unwanted states is
0.016.

\section{Conclusions}

We have demonstrated a scheme for the implementation of quantum
information processing in a system of particles with strong
permanent interactions.  We envisage that a number of these
systems could be used for the design of quantum logical devices.
We have also shown that a system of strongly coupled particles is
qubitless - logical qubits don't directly correspond to any
physical two-state subsystem.  Nevertheless, it is possible to
implement quantum logic operations in this system.

As an example, we presented results on the implementation of the
controlled-NOT gate and Shor's algorithm in a spin-1/2 Heisenberg
chain in a nonuniform magnetic field. The results show that
quantum logic can be implemented with resonant electromagnetic
pulses and that the errors caused by nonresonant transitions can
be effectively reduced.

\begin{acknowledgments}

We are grateful to Tony Leggett for useful discussions. This work
was supported by the Department of Energy under contract
W-7405-ENG-36 and the DOE Office of Basic Energy Sciences, by the
National Security Agency (NSA) and the Advanced Research and
Development Activity (ARDA).
\end{acknowledgments}


\begin{thebibliography}{1}

\bibitem{1}C. Williams, S. Clearwater, {\it Explorations in Quantum Computing}, (Springer-Verlag, Berlin, 1995).

\bibitem{2}G.P. Berman, G.D. Doolen, R. Mainieri,
  V.I. Tsifrinovich, {\it Introduction to Quantum Computers}, (World
  Scientific, Singapore, 1998).

\bibitem{3}{\it Introduction to Quantum Computation and Information}, edited by
H. K. Lo, S.Popescu, T. Spiller, (World Scientific, Singapore,
1998).

\bibitem{a}D.P. DiVincenzo, D. Bacon, J. Kempe, G. Burkard, K.B. Whaley, {\it Nature}, {\bf
408}, 339 (2000).

\bibitem{b}A. Barenco, C.H. Bennett, R. Cleve, D. P. DiVincenzo, N.
  Margolus, P. Shor, T. Sleator, J. Smolin, H. Weinfurter, {\it Phys.
    Rev. A}, {\bf 52}, 3457 (1995).

\bibitem{c}G.P. Berman, G.D. Doolen, G.V. L\'opez, V.I. Tsifrinovich,
{\it Phys. Rev. A}, {\bf 61}, 042307-1 (2000).

\bibitem{e}G.D. Sanders, K.W. Kim, W.C. Holton,
{\it Phys. Rev. A}, {\bf 59}, 1098 (1999).

\bibitem{d}P. Shor, in {\it Proc. 35th Annual Symposium on the
    Foundations of Computer Science}, (IEEE Computer Society Press,
    New York, 1994), p. 124.

\end{thebibliography}
\end{document}